\documentclass[twocolumn,showpacs,preprintnumbers,amsmath,amssymb]{revtex4}

\def\rh{r_{h}}

\def\rc{r_{c}}

\begin{document}
\preprint{gr-qc/0606109}

\title{Thermodynamics of high dimensional Schwarzschild de Sitter spacetimes: variable cosmological constant}
\author{S. Wang}
\email{wjysysgj@163.com} \affiliation{\centerline{College of
Physical Science and Technology,  Central China Normal University,
Wuhan, Hubei, 430079, China} \\
\centerline{and Astrophysics Center, University of Science and
Technology of China, Hefei, Anhui, 230026, China}}

\begin{abstract}
We study the thermodynamic properties of high dimensional
Schwarzschild de Sitter spacetimes with the consideration of quantum
effects. It is shown that by considering the cosmological constant
as a variable state parameter and adding an extra term which denotes
the vacuum energy, both the differential and integral mass formulas
of the first law of Schwarzschild de Sitter spacetimes can be
directly derived from the general Schwarzschild de Sitter metrics in
a simple and natural way. Furthermore, after taking quantum effects
into account, we can see that the cosmological constant must
decrease and the spontaneous decay of the vacuum energy never makes
the entropy of Schwarzschild de Sitter spacetimes decrease. In
addition, though the laws of thermodynamics are very powerful, at
least the third law can not be applied to the Schwarzschild de
Sitter spacetimes. It should be emphasized that these conclusions
come into existence in any dimension.
\end{abstract}


\maketitle
\section{Introduction}
Since the seminal work of Bekenstein \cite{Bekenstein:1973} and
Hawking \cite{Hawking:1974}, black hole thermodynamics
\cite{Bekenstein:1980} had been well established over the past three
decades via a close analogy to the four laws of the usual
thermodynamics system \cite{Bardeen:1973}. In particular, it has
been generally proved \cite{Gauntlett:1999} that for asymptotic flat
black holes in any dimensions, the differential expression of the
first law and the corresponding integral mass formula take the form
\begin{equation}
dM=TdS+\Omega dJ\; ,\label{eq-1}
\end{equation}
\begin{equation}
\frac{D-3}{D-2}M=TS+\Omega J\; , \label{eq-2}
\end{equation}
where $D$ is the spacetime dimensions, $M$, $T$, $S$, $\Omega$ and
$J$ are Arnowitt-Deser-Misner (ADM) mass, temperature,
Benkenstein-Hawking entropy, angular velocity, and  angular momentum
of the black hole, respectively.

Recently, considerable attention has been concentrated on the study
of de Sitter spacetimes and asymptotically de Sitter spacetimes.
This is motivated, basically, by the following two aspects: First,
the Type Ia supernova observations indicate that our universe is in
a phase of accelerating expansion \cite{Perlmutter:1997}. That is to
say, there is a positive cosmological constant in our universe and
our universe might approach to a de Sitter phase in the far future.
Second, defined in a manner analogous to the AdS/CFT correspondence,
an interesting proposal, the so-called dS/CFT correspondence, has
been suggested that there is a dual relation between quantum gravity
on a de Sitter (dS) space and Euclidean conformal field theory (CFT)
on a boundary of de Sitter space
\cite{Witten:2001,Strominger:2001,Hull:1998,Mazur:2001}.

With the advent of AdS/CFT and dS/CFT correspondence, studies of
thermodynamics properties of black holes, especially in the
background geometry with a fixed cosmological constant, have
attracted considerable attention in recent years
\cite{Banados:1992,Dehghani:2002,Wu:2005,Shankaranarayanan:2003,Cai:2001,Gibbons:2005}.
For instance, It has been recently shown by Gibbon \textit{et al}.
\cite{Gibbons:2005} that with a suitable definition of the conserved
charges such as mass and angular momentum, the differential mass
formula (1) keeps to be true for rotating black holes with an
invariable cosmological constant background in four, five and higher
dimensions. However, under the assumption of a fixed cosmological
constant, it is clear that the integral expression (2) can never be
satisfied any more.

In order to rectify this situation, in a previous paper
\cite{Wang:2006}, we have investigated the thermodynamics of the
(2+1)-dimensional BTZ black holes and Kerr-de Sitter spacetimes, and
shown that it is possible for us to consider the cosmological
constant $\Lambda= (D-1)(D-2)/2l^2$ as a varying parameter and
promoting it to a thermodynamics state variable. With this
assumption, both the differential and integral mass formulas of Kerr
de Sitter spacetimes can be satisfied.

As well known, the cosmological constant was originally introduced
by Einstein to keep our universe as a static universe. Though the
discovery by Hubble that the universe is expanding eliminated the
empirical need for a static world model, the disappearance of the
original motivation for introducing the cosmological constant did
not change its status as a legitimate addition to the gravitational
field equations, or as a parameter to be constrained by observation
\cite{Krauss:1995,Cohn:1998,Carroll:2001}. There are two
cosmological constant problems. the old one is to understand why
there is a huge gap between the theoretical and observational values
of the cosmological constant. The new one, which had not been
realized until Type Ia supernova observations provided the evidence
that our universe is accelerating, is to understand why the energy
density of vacuum has the same order of magnitude with the present
mass density of the universe \cite{Weinberg:2000,Garriga:2001}. In
order to solve these two intriguing problems, considerable works had
been done
\cite{Garriga:2001,Padmanabhan:2005,Hamed:2000,Picon:2000}. We do
not further discuss these problems here, but only focus on a
question: it is reasonable whether or not for us to treat the
cosmological constant as a variable parameter?

Twenty years ago, Henneaux and Teitelboim have already shown that
the cosmological constant could be treated as a variable parameter
\cite{Henneaux:1985} (For more details, see Ref.
\cite{Gomberoff:2004}). Before long, It was also conjectured by
Weinberg that the role of the cosmological constant is played by a
slowly varying potential of a scalar field \cite{Weinberg:1987}. the
original motivation for considering cosmological constant as a
variable parameter is the factual need of solving those two
cosmological constant problems that mentioned above
\cite{Weinberg:2000,Garriga:2001}. To my present knowledge, there
are at least three particle physics models, with a discrete spectrum
of vacuum energy density $\rho_{\Lambda}$, would allow cosmological
constant $\Lambda$ to take different values. The first
discrete-$\rho_{\Lambda}$ model, introduced by Abbott
\cite{Abbot:1987}, considers a self-interacting scalar field $\phi$
with a washboard potential $V(\phi)$. In this model, the vacuum at
$\phi=\phi_{n}$ has energy density $\rho_{\Lambda
n}=n\epsilon+const$, and can decay through bubble nucleation to the
vacuum at $\phi_{n-1}$. The upward quantum jumps from $\phi_{n-1}$
to $\phi_{n}$ are also possible \cite{Lee:1987}. The second
discrete-$\rho_{\Lambda}$ model, first discussed by Brown and
Teitelboim \cite{Brown:1987}, assumes that the cosmological constant
is due to a four-form field
$F^{\alpha\beta\gamma\delta}={F\over{\sqrt{-g}}}\epsilon^{\alpha\beta\gamma
\delta}$, which can change its value through the nucleation of
branes. The total vacuum energy density is given by
$\rho_{\Lambda}=\rho_{bare}+{1\over{2}}F^2$, where $\rho_{bare}<0$
is the `bare' cosmological constant at $F=0$; The change of the
field $\Delta F=\pm q$, where $q=const$ is fixed by the model (For
related work, see Ref. \cite{Feng:2001} ). The third
discrete-$\rho_{\Lambda}$ model, developed by Bousso and Polchinski
\cite{Bousso:2000}, is a different version of the four-form model.
In this model, several four-forms fields $F_i$ are present, so the
total vacuum energy density takes the form
$\rho_{\Lambda}=\rho_{bare}+{1\over{2}}\sum_i F_i^2$. It is clear
that these four-form fields can also change their value through the
nucleation of branes (It should be pointed out that this model is
quite different from the second model. For more detail, see Refs.
\cite{Bousso:2000,Banks:2001} ). As Dvali and Vilenkin argued, the
varying of four-form field can also be attributed to a spontaneously
broken discrete symmetry \cite{Dvali:2001}. Since the vacuum energy
density is a variable parameter, as a measure of vacuum energy
density \cite{Carroll:2001,Garriga:2001}, the cosmological constant
should not be viewed as a fixed parameter, also. This assumption of
variable cosmological constant is also supported by some other works
\cite{Creighton:1995,Caldarelli:2000,Shapiro:2003,Horvat:2004,Sekiwa:2006}.

Recently, with the assumption of variable cosmological constant, the
first and second law of four-dimensional Kerr-Newman black hole in
de Sitter backgrounds had already been well studied
\cite{Sekiwa:2006}. However, to my present knowledge, the
thermodynamic properties of higher dimensional case in the
background geometry with a variable cosmological constant have not
been investigated. In the present paper, we will expand the
investigation to the higher dimensional case. This subject is mainly
motivated by the fact that recent brane-world scenarios
\cite{Randall:1999} predict the emergence of a TeV-scale gravity in
the higher-dimensional theories, thus open the possibility to
explore extra dimensions by making tiny black holes in the
high-energy collides \cite{Dimopoulos:2001} (LHC) that to be running
soon. Within this context, it is not unnecessary to take higher
dimensional case into account. We will show that the differential
and integral mass formulas of first law take the forms
\begin{equation}
dM=TdS+\Theta dl\; ,\label{eq-3}
\end{equation}
\begin{equation}
\frac{D-3}{D-2}M=TS+\frac{1}{D-2}\Theta l \; ,\label{eq-4}
\end{equation}
Where $\Theta$ is a generalized force conjugate to the cosmological
radius $l$; the combination $\Theta l$ has the dimension of energy
and denotes the vacuum energy. It will be proved that both the mass
formulas (3) and (4) can be directly derived from the general
Schwarzschild de Sitter metrics, and come into existence in any
dimension.

Recently, It has been pointed that when one consider the second law
of black hole thermodynamics, the quantum effects must be taken into
account \cite{Sekiwa:2006}. It is found that after taking quantum
effects into account, the cosmological constant must decrease and
the spontaneous decay of the vacuum energy never makes the entropy
of the Schwarzschild de Sitter spacetimes decrease.

The authors of Ref. \cite{Huang:2002} argue that all the four laws
of thermodynamics can be applied to the Schwarzschild de Sitter
spacetimes. However, it seems that though the laws of thermodynamics
are very powerful, at least the third law can not be applied to the
Schwarzschild de Sitter spacetimes.

The organization of this paper is as follows: In Sec \ref{section2},
we study the first law of black hole thermodynamics in Schwarzschild
de Sitter spacetimes. With the assumption of variable cosmological
constant, we can see that both the mass formulas (3) and (4) can be
directly derived from the general Schwarzschild de Sitter metrics,
and come into existence in any dimension. In Sec \ref{section3}, we
discuss the second law of Schwarzschild de Sitter spacetimes. It is
found that after taking quantum effects into account, the
cosmological constant must decrease and the spontaneous decay of the
vacuum energy never makes the entropy of the Schwarzschild de Sitter
spacetimes decrease. In Sec \ref{section4}, we study other laws of
thermodynamics and find that the third law can not be applied to the
Schwarzschild de Sitter spacetimes. Sec \ref{section5} contains a
brief remark.

Throughout this paper, the metric signature adopted is $(-,+,+,+)$,
and the natural units $c=\hslash=G=\kappa=1$ are used.
\section{THE FIRST LAW}\label{section2}
It has been proved that in Boyer-Linquist coordinates, the general
Schwarzschild de Sitter metrics for arbitrary dimension takes the
form \cite{Gibbons:2005}
\begin{eqnarray}
ds^{2} &=& -\Big(1-\frac{r^{2}}{l^{2}}\Big)d\tau^{2}
+\frac{2m}{U}d\tau^{2}+\frac{Udr^{2}}{V-2m} \nonumber \\
&& +\sum\limits_{i=1}^N r^{2}\big(d\mu_{i}^{2}+\mu_{i}^{2}d\varphi_{i}^{2}\big) \nonumber \\
&& +\frac{l^{-2}}{1-r^{2}l^{-2}} \Big(\sum\limits_{i=1}^N
r^{2}\mu_{i}d\mu_{i}\Big)^{2}\; ,\label{eq-5}
\end{eqnarray}
Where
\begin{equation}
U=r^{D-3}\; , \qquad V=r^{D-3}\big(1-\frac{r^{2}}{l^{2}}\big)\;
.\label{eq-6}
\end{equation}
This metric depends on two parameters: the mass $m$, and the
cosmological radius $l$, which is related to the cosmological
constant $\Lambda$ by $\Lambda= (D-1)(D-2)/2l^2$.

The horizons are located at
\begin{equation}
r^{D-3}\big(1-\frac{r^{2}}{l^{2}}\big)-2m=0\; .\label{eq-7}
\end{equation}
This algebra equation has several positive solutions, which
associated with multiple horizons. The largest one denotes the
cosmological horizon radius $r_{c}$, and the smallest one
corresponds to the black hole horizon radius $r_{h}$. The relation
between the physical mass and the geometrical parameter can be
written as \cite{Gibbons:2005}
\begin{equation}
M_{h}=-M_{c}=f(D)m\; ,\label{eq-8}
\end{equation}
where $M_{c}$ and $M_{h}$ are mass of cosmological horizon and black
hole horizon respectively, $f(D)$ is a function of the spacetime
dimensions $D$. It can be seen as follows, the expressions of the
differential and integral mass formulas of the first law have
nothing to do with the form of $f(D)$.

It is well known that the Hawking temperatures associated with the
cosmological horizon and black hole horizon, respectively, are not
equal \cite{Gibbons:1977}. They both emit Hawking radiation at the
corresponding temperatures and are not in thermal equilibrium. As
Teitelboim pointed out \cite{Teitelboim:2002}, we must take into
account two different idealized physical systems. One is a black
hole horizon enclosed in a cosmological boundary, the other is a
cosmological horizon enclosed in a black hole boundary. When we
discuses the thermodynamics properties of either one of them, the
other one must be viewed as a boundary. (Owing to the existence of
multiple horizons, the thermodynamic aspects of Schwarzschild de
Sitter spacetimes have not been well explored so far. Recently, it
has been argued that in general for a spherically symmetric
spacetime with multiple horizons, the temperature of radiation is
proportional to the effective surface gravity, and the equilibrium
temperature in Schwarzschild de Sitter spacetimes is the harmonic
mean of cosmological and black hole horizon temperatures. For more
detail, see Ref. \cite{Shankaranarayanan:2003} )

Let us discuss the cosmological horizon first. As mentioned above,
the black hole horizon must be treated as a boundary. The area of
the cosmological horizon is
\begin{equation}
A_{c}=\frac{(D-1)\pi^{(D-1)/2}}{\Gamma[(D+1)/2]}r_{c}^{D-2}\;
,\label{eq-9}
\end{equation}
Note that the macroscopic entropy-area law which relates
thermodynamic entropy to the area of event horizon is universally
valid for any types of black holes belonging to Schwarzschild
family, thus
\begin{equation}
S_{c}=\frac{A_{c}}{4}=\frac{(D-1)\pi^{(D-1)/2}}{4\Gamma[(D+1)/2]}r_{c}^{D-2}=\frac{r_{c}^{D-2}}{g(D)}\;
,\label{eq-10}
\end{equation}
Where $g(D)$ is a function of the spacetime dimensions $D$. Making
use of eps. (7), (8) and (10), one can obtain
\begin{equation}
M_{c}=-\frac{f(D)}{2}[g(D)S_{c}]^{\frac{D-3}{D-2}}+\frac{f(D)}{2l^{2}}[g(D)S_{c}]^{\frac{D-1}{D-2}}\;
,\label{eq-11}
\end{equation}
This is the so-called generalized Smarr formula \cite{Smarr:1973},
and it contains all the information about the thermodynamics state
of cosmological horizon.

Just like our universe should be treated as an open system
\cite{Prigogine:1989}, de Sitter spacetimes should not be regarded
as an isolated system \cite{Huang:2002}. This is because when the
cosmological constant changes, the horizon shrinks or expands and
thus sweeps some region with a nonzero cosmological constant,
leading to the variation of the four-volume of the Euclidean de
Sitter spacetimes.

As mentioned above, the cosmological constant should be treated as a
variable state parameter. It must be emphasized that this assumption
can never be justified in classical theory (i.e. general relativity)
\cite{Sekiwa:2006}. If we want to explain it on physical grounds, we
must take quantum effects into account. In fact, particle physics
has already provided us several approaches to explain why should
cosmological constant $\Lambda$ be a variable. For instance, it can
be well explained by the four-form model
\cite{Brown:1987,Feng:2001}. This model has recently attracted much
attention because four-form fields with appropriate couplings to
branes naturally arise in the context of M-theory
\cite{Garriga:2001}.

The author of Ref. \cite{Sekiwa:2006} argues that one should choose
$\Lambda$ as a thermodynamics state variable for a clearer physical
meaning. However, it is not absolutely necessary for one to do so.
For instance, one can choose $l^{n}$ (where $n\leq D$) as a
thermodynamics state variable. When $n=1$, $l$ has dimension of
length and $\Theta$ can be treated as generalized force; when $n=2$,
$l^{2}$ has dimension of area and $\Theta$ can be treated as
generalized surface gravity; when $n=3$, $l^{3}$ has dimension of
volume and $\Theta$ can be treated as generalized pressure; when
$n=-2$, $l^{-2}$ has the same dimension with cosmological constant
and $\Theta$ can be interpreted as generalized volume (see
Ref.\cite{Sekiwa:2006} ). If only the combination $\Theta l^{n}$ has
the dimension of energy and denotes the vacuum energy (It must be
emphasized that this vacuum energy term can be provided by quantum
effects. For related work, see Refs.
\cite{Bousso:1998,Odintsov:1999,Padmanabhan:2002} ), there is no
essential difference among the different choices of state variable.
For simplicity, here we still choose $l$ as thermodynamics state
variable.

The Hawking temperature and the generalized force can be computed
\begin{equation}
T_{c}=-\frac{f(D)g(D)[(D-3)l^{2}-(D-1)r_{c}^{2}]}{2(D-2)r_{c}l^{2}}\;
,\label{eq-12}
\end{equation}
\begin{equation}
\Theta_{c}=-\frac{f(D)}{l^{3}}[g(D)S_{c}]^{\frac{D-1}{D-2}}=-\frac{f(D)r_{c}^{D-1}}{l^{3}}\;
,\label{eq-13}
\end{equation}
Then one can obtain
\begin{equation}
dM_{c}=T_{c}dS_{c}+\Theta_{c} dl\; ,\label{eq-14}
\end{equation}
\begin{equation}
\frac{D-3}{D-2}M_{c}=T_{c}S_{c}+\frac{1}{D-2}\Theta_{c} l \;
,\label{eq-15}
\end{equation}
We can see that both the differential and integral expression can
satisfy the formulas (3) and (4). It is clear that if we do not
consider the cosmological constant as a variable state parameter,
the integral mass formulas (15) can never be obtained.

Next, we will discuss the black hole horizon. Then the cosmological
horizon must be treated as a boundary. Obviously, there exists a
situation analogous to that of electric charge on a two-sphere
\cite{Teitelboim:2002}. In that case, if a charge $q$ is placed at
the North Pole, an opposite charge, $-q$, must appear at the South
Pole. The same phenomenon occurs here for the energy in $r-t$ space.
The area of the black hole horizon is
\begin{equation}
A_{h}=\frac{(D-1)\pi^{(D-1)/2}}{\Gamma[(D+1)/2]}r_{h}^{D-2}\;
,\label{eq-16}
\end{equation}
According to the Bekenstein-Hawking entropy-area law
\begin{equation}
S_{h}=\frac{A_{h}}{4}=\frac{(D-1)\pi^{(D-1)/2}}{4\Gamma[(D+1)/2]}r_{h}^{D-2}=\frac{r_{h}^{D-2}}{g(D)}\;
,\label{eq-17}
\end{equation}
Making use of eps. (7), (8) and (17), one can obtain
\begin{equation}
M_{h}=\frac{f(D)}{2}[g(D)S_{h}]^{\frac{D-3}{D-2}}-\frac{f(D)}{2l^{2}}[g(D)S_{h}]^{\frac{D-1}{D-2}}\;
,\label{eq-18}
\end{equation}

In fact, there exist other methods to calculate the conserved
quantities in de Sitter spacetimes. For example, in Ref.
\cite{Cai:2002}, the authors use the Balasubramanian-Boer-Minic
(BBM) prescription \cite{Balasubramanian:2002} to calculate the
conserved quantities for cosmological horizon and the Abbott-Deser
(AD) prescription \cite{Abbott:1982} to calculate the ones for black
hole horizon, respectively. However, if one uses these methods, the
generalized Smarr formula [(11) and (18)] can not be obtained, and
then the integral mass formulas (4) can never be derived.

The Hawking temperature and the generalized force can be computed
\begin{equation}
T_{h}=\frac{f(D)g(D)[(D-3)l^{2}-(D-1)r_{h}^{2}]}{2(D-2)r_{h}l^{2}}\;
,\label{eq-19}
\end{equation}
\begin{equation}
\Theta_{h}=\frac{f(D)}{l^{3}}[g(D)S_{h}]^{\frac{D-1}{D-2}}=\frac{f(D)r_{h}^{D-1}}{l^{3}}\;
,\label{eq-20}
\end{equation}
Then one can obtain
\begin{equation}
dM_{h}=T_{h}dS_{h}+\Theta_{h} dl\; ,\label{eq-21}
\end{equation}
\begin{equation}
\frac{D-3}{D-2}M_{h}=T_{h}S_{h}+\frac{1}{D-2}\Theta_{h} l \;
,\label{eq-22}
\end{equation}
We have succeeded in deriving the differential and integral mass
formulas of the first law from the general Schwarzschild de Sitter
metrics in a simple and natural way. It is clear that these two
formulas come into existence in any dimension. As Carlip and Vaidya
pointed out \cite{Carlip:2003}, one should take into account not
just the black hole, but its surroundings as well. It is interesting
to further extend the present investigation to the full thermal
environment of a black hole.
\section{THE SECOND LAW}\label{section3}
Let us turn our attention to the second law of thermodynamics. As
mentioned above, the quantum effects (i.e. quantum evaporate and
quantum anti-evaporate process. For more detail, see Refs.
\cite{Bousso:1998,Odintsov:1999} ) must be taken into account.

For black hole horizon,
\begin{equation}
-T_{h}dS_{h}=\frac{f(D)r_{h}^{D-1}}{l^{3}}dl\; ,\label{eq-23}
\end{equation}
The left side denotes the entropy loss of black hole horizon, and
the right side denotes the decrease of vacuum energy inside the
black hole horizon. We can see that the decrease of vacuum energy is
equal to the entropy decreasing of black hole horizon. For observer
outside the black hole horizon, the total energy variation is seen
as the decrease of entropy inside the black hole horizon. This
energy is carried away from inside to outside by means of Hawking
radiation.

For cosmological horizon,
\begin{equation}
T_{c}dS_{c}=\frac{f(D)r_{c}^{D-1}}{l^{3}}dl\; ,\label{eq-24}
\end{equation}
The left side expresses the entropy increase of cosmological
horizon, and the right side expresses the increase of vacuum energy
in the visible region $r_{h}<r<r_{c}$. Vacuum energy is transformed
quantum mechanically to the energy of radiation from outside to
inside. That is to say, the origin of Hawking radiation can be
attributed to the varying of cosmological constant
\cite{Sekiwa:2006}. Noting that $\Lambda\propto l^{-2}$, it is easy
for us to see that the generalized entropy $S_{c}$ increases if and
only if the cosmological constant $\Lambda$ decreases. In other
words, the cosmological constant must decreases through quantum
effects owing to the generalized second law requires that
generalized entropy never decreases for all physical processes.

Base on eqs. (23) and (24), one can obtain
\begin{equation}
T_{c}dS_{c}=-T_{h}dS_{h}+\frac{r_{c}^{D-1}-r_{h}^{D-1}}{l^{3}}f(D)dl\;
,\label{eq-25}
\end{equation}
This equation implies that the entropy increase of cosmological
horizon is due to the thermal radiation from the cosmological and
black hole horizon. That is to say, the origins of entropy increase
are the increase of vacuum energy in the visible region
$r_{h}<r<r_{c}$ (because $\rc$ increases and $\rh$ shrinks when
cosmological constant $\Lambda$ decreases, the total vacuum energy
in the region $\rh < r < \rc$ must increases) and Hawking radiation
from these two horizons. Since the generalized second law requires
that generalized entropy never decreases for all physical processes,
the cosmological constant must decrease. In other words, the
spontaneous decay of the vacuum energy never makes the entropy of
the Schwarzschild de Sitter spacetimes decrease.

An example of the spontaneous decay of the vacuum energy is the
quantum tunneling from the false vacuum of Higgs potential
\cite{Huang:2002}. The final potential in the quantum tunneling is
never higher than the initial potential, which corresponds to
monotonically decrease of the cosmological constant. (Indeed, since
Parikh and Wilzcek \cite{Parikh:2000} presented a greatly simplified
model to implement the Hawking radiation as a semi-classical
tunnelling process, considerable attention has been focused on
extending this semi-classical tunnelling method to various cases of
black holes. For more detail, see Refs.
\cite{Vagenas:2001,Hemming:2001,Wu:2006} )
\section{THE ZEROTH AND THIRD LAWS}\label{section4}
The authors of Ref. \cite{Huang:2002} argue that all the four laws
of thermodynamics can be applied to the Schwarzschild de Sitter
spacetimes (universes). Except for the first law and the second law
we have discussed above, other laws can be written as:

The zeroth law: the cosmological constant should be same at all
places.

The third law: the cosmological constant can never reach to zero by
finite physical processes.

It is obvious that the zeroth law is undoubtedly correct. Since the
cosmological constant is independent of spacetime coordinates, if it
decreases inside the cosmological horizon, it must decrease at the
outside also. That is to say, the vacuum energy density should
decrease at same velocity everywhere. Therefore, though cosmological
constant must decay, we can still say that it should be same at all
places. However, the third law seems not so reasonable.

As well known, the third law of ordinary thermodynamics system
demands $\lim\limits_{\Lambda \to 0} S=0$. We have already obtained
\begin{equation}
S=\frac{(D-1)\pi^{(D-1)/2}}{4\Gamma[(D+1)/2]}r^{D-2}\;
,\label{eq-26}
\end{equation}
It is clear that the cosmological constant $\Lambda$ (or the
cosmological radius $l$) has nothing to do with entropy $S$, i.e.
$\lim\limits_{\Lambda \to 0} S\neq 0$. In addition, base on the above discussions
\begin{equation}
T=\frac{f(D)g(D)[(D-3)l^{2}-(D-1)r^{2}]}{2(D-2)rl^{2}}\;
,\label{eq-27}
\end{equation}
We can see that in the limit $l\rightarrow\infty$ (i.e.
$\Lambda\rightarrow0$), it is clear that $T\neq0$. So there is no
inherent relation between temperature $T$ and cosmological constant
$\Lambda$. Base on these discussions, we can say that there is no
enough evidence to support the third law of de Sitter spacetimes
from the view of mathematics.

Let us turn to the view of physics. Though there is a nonzero
positive cosmological constant in our present universe, It is
extremely small and will decay all the times. So far, there is no
principle that could stop this process of decaying being discovered.
Therefore, the third law of thermodynamics cannot be directly
applied to the Schwarzschild de Sitter spacetimes.
\section{Concluding remarks}\label{section5}
In summary, we have investigated the thermodynamic properties of
high dimensional Schwarzschild de Sitter spacetimes. By treating
cosmological constant as a variable state parameter and adding an
extra term that denotes the vacuum energy, we can prove that both
the differential and integral mass formulas [(3) and (4)] of the
first law of Schwarzschild de Sitter spacetimes can be directly
derived from the general Schwarzschild de Sitter metrics.
Furthermore, it is shown that the cosmological constant must
decrease and the spontaneous decay of the vacuum never makes the
entropy of the Schwarzschild de Sitter universe decrease. In
addition, it is found that though the laws of thermodynamics are
very powerful, at least the third law can not be applied to the
Schwarzschild de Sitter spacetimes. It is obvious that these
conclusions come into existence in any dimension.

Since the Type Ia supernova observations indicate that our universe
is in a phase of accelerating expansion, recently there is much
attention \cite{Cai:2005,Wang:2007} have being paid to investigate
the thermodynamic properties of the universe with a positive
cosmological constant. It should be pointed that these works are
base on the background geometry with a fixed cosmological constant.
It is interesting to investigate the Thermodynamics properties of an
accelerated expanding universe with the framework of variable
cosmological constant. This issue deserves further research in the
future.
\begin{acknowledgments}
I am grateful to Prof. S.Q. Wu for useful and valuable discussions.
I also thank S. Odintsov, T. Padmanabhan and S. Shankaranarayanan
for helpful suggestions.
\end{acknowledgments}


\end{document}